\documentclass{article}
\usepackage{spconf,amsmath,graphicx}
\usepackage{amsfonts,amssymb}
\usepackage{booktabs}
\usepackage{multirow}
\pdfoutput=1

\title{Gaussian-smoothed imbalance data improves speech emotion recognition}
\name{Xuefeng Liang$^{\star \dagger}$ \qquad Hexin Jiang$^{\star}$ \qquad Wenxin Xu$^{\star}$ \qquad Ying Zhou$^{\star}$}
\address{ $^{\star}$ School of Artificial Intelligence, Xidian University, China
\\ $^{\dagger}$Pazhou Lab, Huangpu, China}
\begin{document}
%
\maketitle
\begin{abstract}
In speech emotion recognition tasks, models learn emotional representations from datasets. We find the data distribution in the IEMOCAP dataset is very imbalanced, which may harm models to learn a better representation. To address this issue, we propose a novel Pairwise-emotion Data Distribution Smoothing (PDDS) method. PDDS considers that the distribution of emotional data should be smooth in reality, then applies Gaussian smoothing to emotion-pairs for constructing a new training set with a smoother distribution. The required new data are complemented using the mixup augmentation. As PDDS is model and modality agnostic, it is evaluated with three SOTA models on the IEMOCAP dataset. The experimental results show that these models are improved by 0.2\% $\sim$ 4.8\% and 1.5\% $\sim$ 5.9\% in terms of WA and UA. In addition, an ablation study demonstrates that the key advantage of PDDS is the reasonable data distribution rather than a simple data augmentation.
\end{abstract}
\section{Introduction}
\label{sec:intro}

\begin{figure}[htb]
\centering
\centerline{\includegraphics[width=8.5cm]{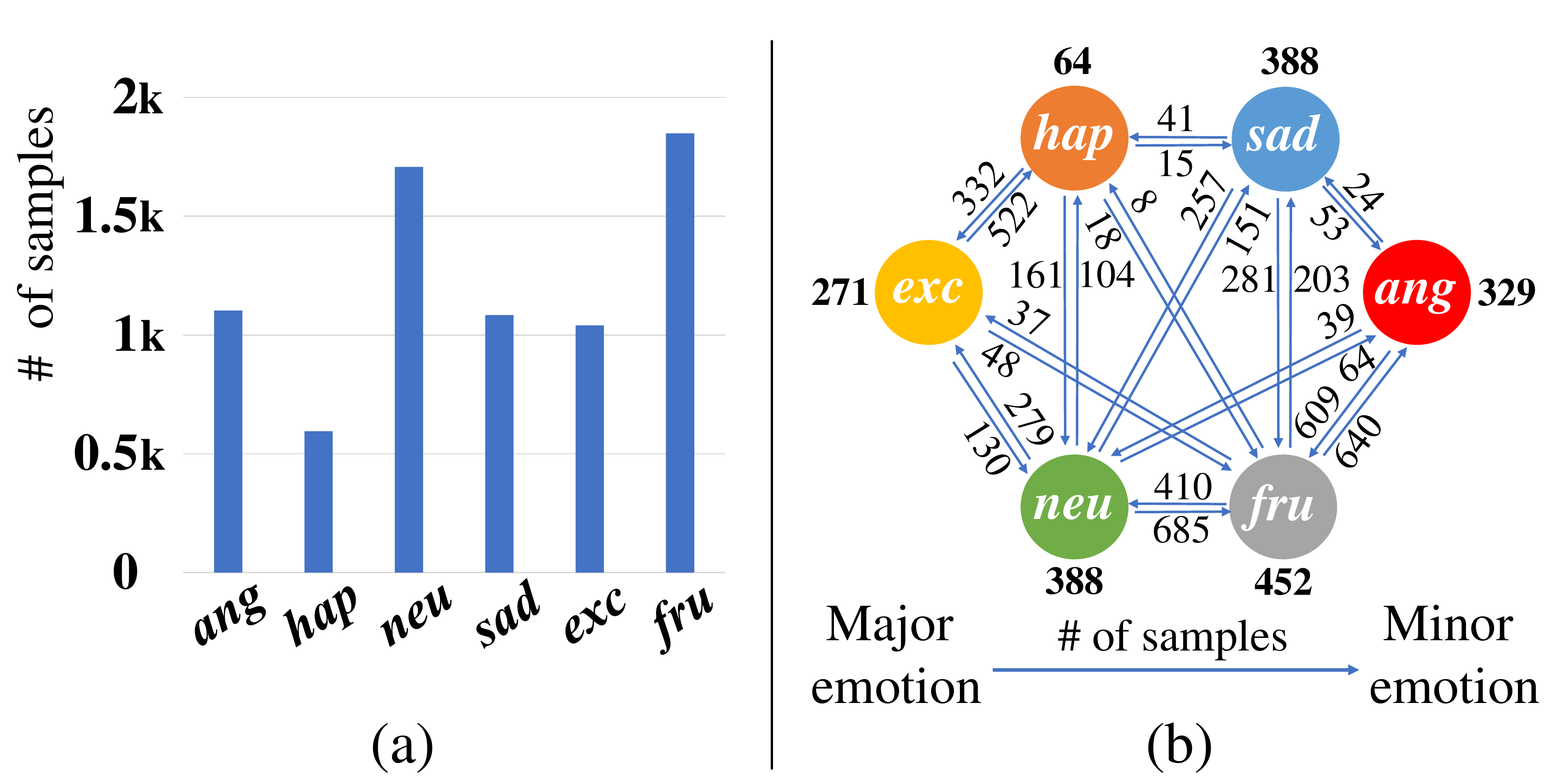}}
\caption{(a) Data distribution in the IEMOCAP when the label is one-hot. (b) The statistics of clear and ambiguous samples in the IEMOCAP. Each vertex represents an emotion category with only clear samples. The bold number is the quantity of clear samples. Each directed edge denotes a set of ambiguous samples, whose tail is the major emotion and head is the minor emotion. The quantity of ambiguous samples is on the edge.}
\label{fig:1}
\end{figure}

Speech emotion recognition (SER) is of great significance to understanding human communication. SER techniques have been applied to many fields, such as video understanding \cite{gao2021pairwise}, human-computer interaction \cite{cowie2001emotion}, mobile services \cite{huahu2010application} and call centers \cite{gupta2007two}. Up to now, plenty of deep learning based SER methods have been proposed \cite{parry2019analysis, poria2017context}. Recently, multimodal emotion recognition attracted more attention \cite{atmaja2019speech, delbrouck-etal-2020-modulated, sun2021multimodal} because of its richer representation and better performance. Most of these studies assigned one-hot labels to utterances. In fact, experts often have inconsistent cognition of emotional data, thus, the one-hot label is obtained by majority voting. Figure \ref{fig:1}(a) shows the emotional data distribution in the IEMOCAP dataset \cite{busso2008iemocap}.

Later, some studies argued that the one-hot label might not represent emotions well. They either addressed this problem through multi-task learning \cite{Lotfian2018}, soft-label \cite{fayek2016modeling, ando2018soft} and multi-label \cite{Ando2019}, or enhanced the model’s learning ability of ambiguous emotions by dynamic label correction \cite{Fujioka2020} and label reconstruction through interactive learning \cite{zhou2022multi}. These works usually defined the samples with consistent
experts’ votes as clear samples and those with inconsistent votes as ambiguous samples. The statistics of clear and ambiguous samples in the IEMOCAP dataset are shown in Fig. \ref{fig:1}(b). After statistical analysis, we find that the data distribution is imbalanced, especially for ambiguous samples. For example, the quantity of ambiguous samples between \textit{anger} and \textit{frustration} is abundant, while that between \textit{anger} and \textit{sadness} is very few. Meanwhile, the quantity of clear samples of \textit{happiness} counts much less than other clear emotion categories. We consider such distribution is unreasonable that may prevent the model from learning a better emotional representation. The possible reason is that the votes come from few experts, which is a rather sparse sampling of human population.
\begin{figure}[htb]
\centering
\centerline{\includegraphics[width=8.5cm]{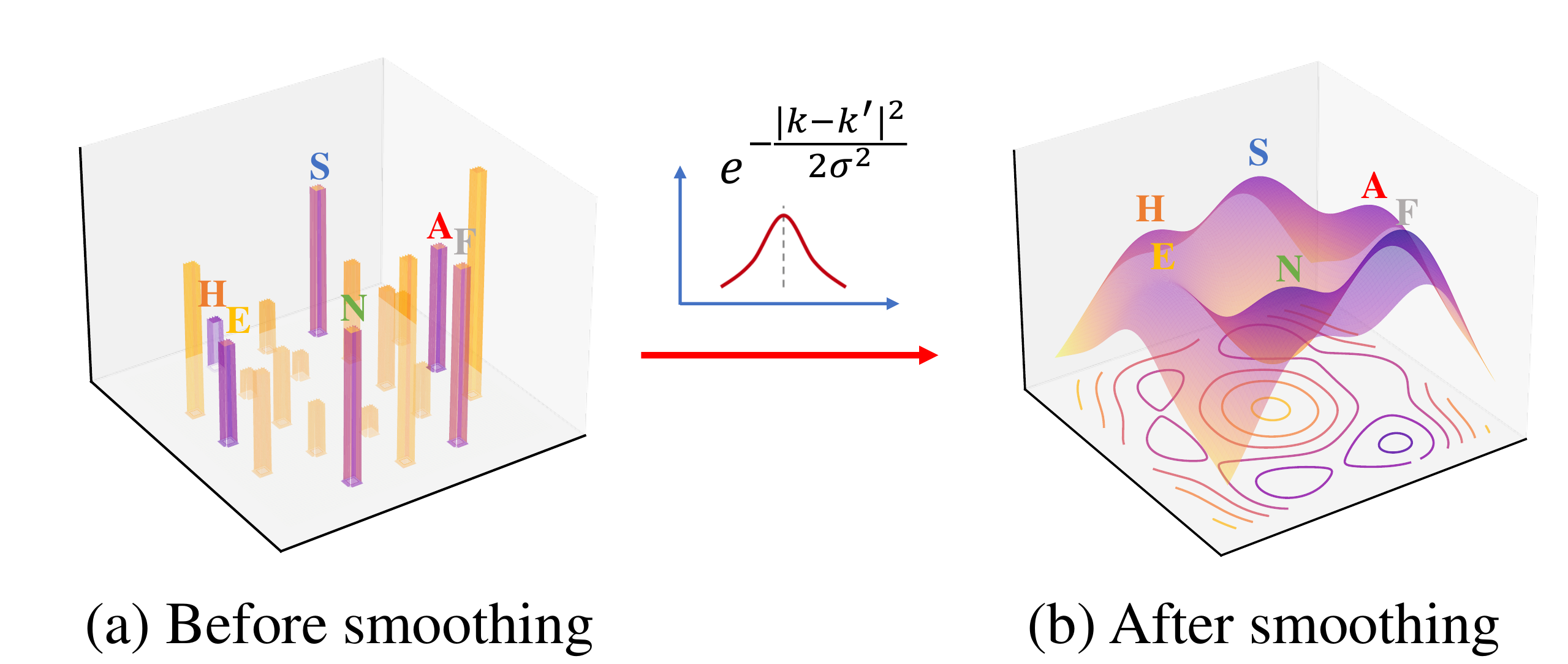}}
\caption{The data distributions before and after smoothing. The purple bars represent clear samples and the orange bars represent ambiguous samples.}
\label{fig:2}
\end{figure}
\begin{figure}[htb]
\centering
\centerline{\includegraphics[width=8.5cm]{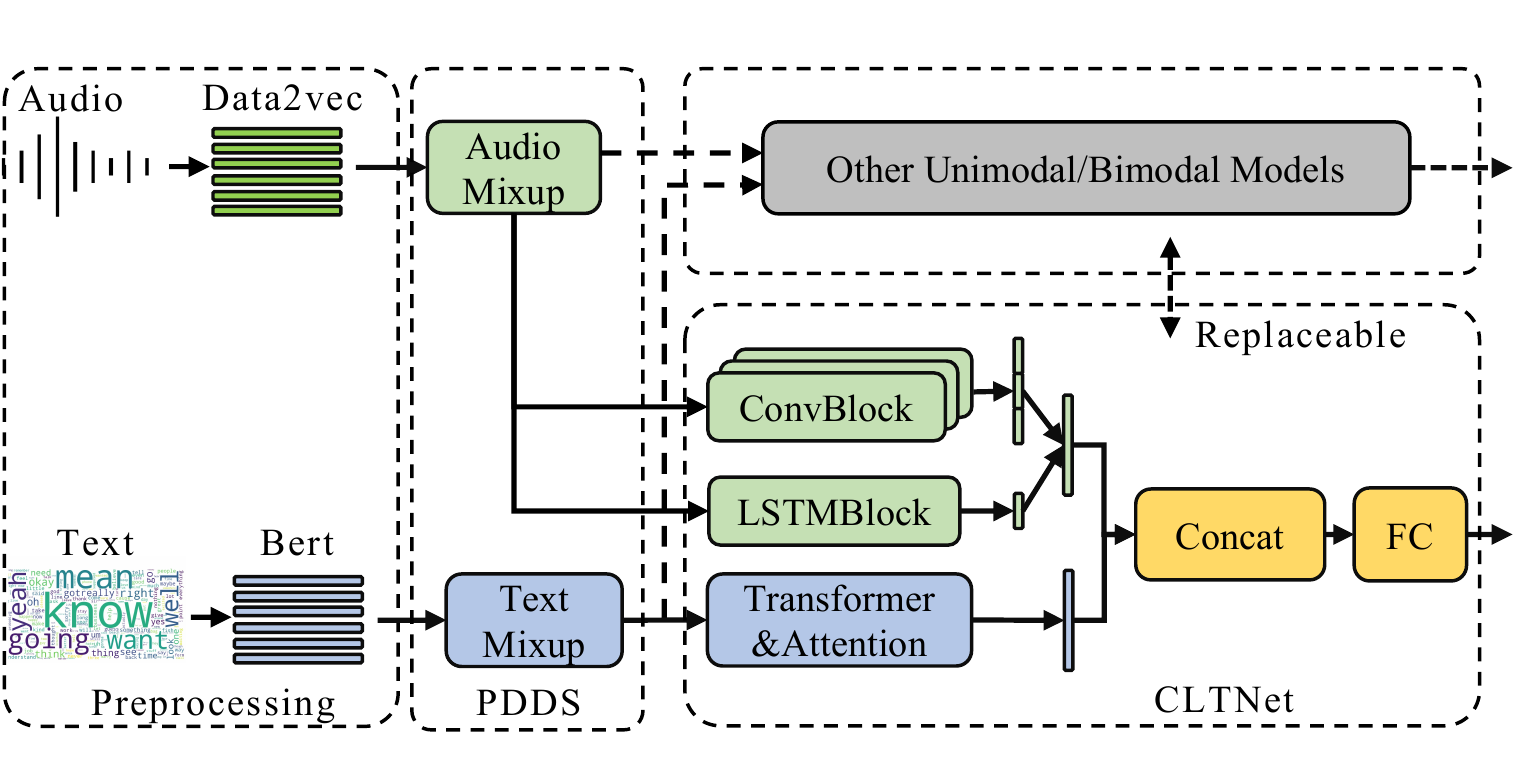}}
\caption{The framework of our approach. It consists of three modules: (1) Data preprocessing; (2) PDDS; (3) CLTNet.}
\label{fig:3}
\end{figure}

We think that clear and ambiguous emotional data shall follow a smooth statistical distribution in the real world. Based on this assumption, we propose the Pairwise-emotion Data Distribution Smoothing (PDDS) to address the problem of unreasonable data distribution. PDDS applies Gaussian smoothing on the data distribution between clear emotion-pairs, which augments ambiguous samples up to reasonable quantities, meanwhile balances the quantities of clear samples in all categories to be close to each other. Figure \ref{fig:2} shows the data distributions before and after smoothing. To complement the missing data, we use a feature-level mixup between clear samples to augment the data.

As PDDS is model and modality agnostic, we evaluate it on three SOTA methods, as well as the unimodal and bimodal data on the IEMOCAP dataset. The results show that these models are improved by 0.2\% $\sim$ 4.8\% and 1.5\% $\sim$ 5.9\% in terms of WA and UA. Our proposed CLTNet achieves the best performance. The ablation study reveals that the nature of superior performance of PDDS is the reasonable data distribution rather than simply increasing the data size.

\section{Method}
\label{sec:method}

\begin{figure}[htb]
\centering
\centerline{\includegraphics[width=8.0cm]{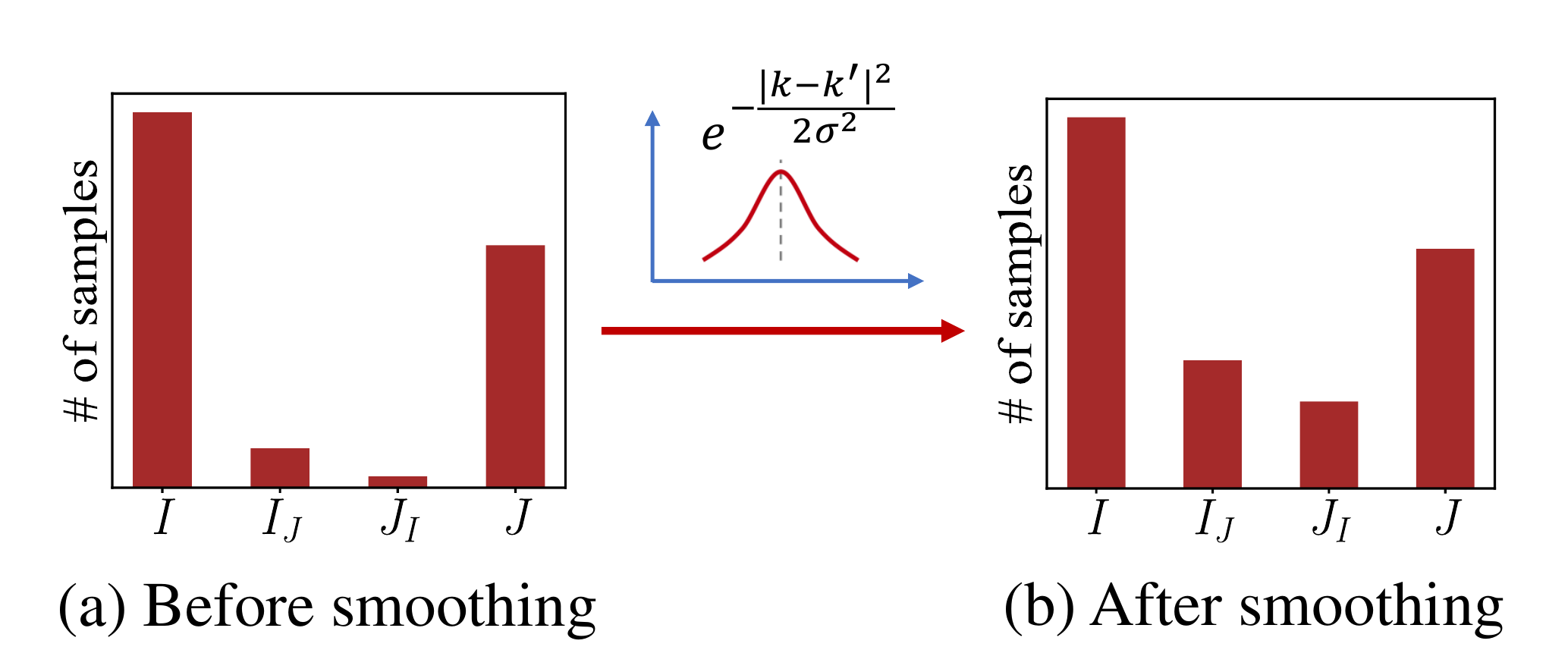}}
\caption{An example of smoothing the quantity distribution of an emotion-pair.}
\label{fig:4}
\end{figure}

Figure \ref{fig:3} shows that our proposed framework includes three modules: (1) Preprocessing module. It extracts audio features using the pre-trained data2vec \cite{pmlr-v162-baevski22a} and text features using the pre-trained Bert \cite{devlin-etal-2019-bert}. (2) Pairwise-emotion Data Distribution Smoothing module (PDDS). It smooths the unreasonable data distribution, and is a plug-in module that can be applied to other SER methods. (3) CLTNet: a proposed model for utterance-level multimodal emotion recognition.

\subsection{Pairwise Data Distribution Smoothing}
\label{ssec:pdds}

\subsubsection{Quantity smoothing}
\label{sssec:qs}

Suppose there are $c$ emotions in the dataset. As the label of an ambiguous sample often comes from two distinct emotions, we construct a clear-ambiguous-clear distribution for the population of four types of samples between every two emotions $i$ and $j$, where $i, j\in \{1,…,c \}, i\neq j$. They are \textit{clear samples} $I$ only containing emotion $i$, \textit{ambiguous samples} $I_J$ containing major emotion $i$ and minor emotion $j$, \textit{ambiguous samples} $J_I$ containing major emotion $j$ and minor emotion $i$, and \textit{clear samples} $J$ only containing emotion $j$, as shown in Fig. \ref{fig:4}.

 We think that the quantity distribution of these four types of samples in every emotion-pair should be statistically smooth, so a Gaussian kernel is convolved with the distribution to have a smoothed version,
\begin{equation}
n_k=\sum_{k' \in K} e^{-\frac{\|k-{k}' \|^2}{2\sigma^2}}n_{{k}'},
\label{eq:1}
\end{equation}
where $K= \{k-1, k, k+1\}$ denotes the indexes of $\{I, I_J, J_I\}$ or $\{ I_J, J_I, J\}$ when $k$ is the index of $I_J$ or $J_I$, $ n_{{k}'}$  is the quantity of samples in type $k$ before smoothing, and $n_k$  is the quantity of samples after smoothing. For clear samples, if their quantity in an emotion category is too small, they are augmented until the quantity reaches that in other categories. The smoothed quantity distribution of all emotion-pairs is shown in Fig. \ref{fig:2}(b).

\subsubsection{Mixup augmentation}
\label{sssec:mixup}

After smoothing the data distribution, the quantities of some ambiguous samples in the original dataset are less than the expectation. To complete data, a feature-level data augmentation, Mixup \cite{zhang2018mixup}, is applied to augment those samples. Mixup can improve the generalization ability of the model by generating new training samples by linearly combining two clear samples and their labels. It follows the rules
\begin{equation}
x_{mix}=p x_\alpha+(1-p)x_\beta,
\label{eq:2}
\end{equation}
\begin{equation}
y_{mix}=p y_\alpha+(1-p)y_\beta,\; p \in \left [ 0,1\right ],
\label{eq:3}
\vspace{0.1cm}
\end{equation}
where $x_\alpha$ and $x_\beta$  are the features from a clear sample of the major emotion and a clear sample of the minor emotion, respectively. $x_{mix}$ is the feature of the new sample. $y_\alpha$ and $y_\beta$ are the one-hot labels of $x_\alpha$ and $x_\beta$, and $y_{mix}$ is the label distribution of the new sample. To avoid undersampling, we use the original data when the original quantity of ambiguous samples has met or exceeded the smooth distribution.

\subsection{CLTNet}
\label{ssec:cltnet}
To further verify the effectiveness of PDDS, we design a simple but effective utterance-level multimodal fusion network, named CLTNet, which uses CNN, LSTM and Transformer to extract multimodal emotional features as shown in Fig. \ref{fig:3}. Firstly, for the audio modality, the acoustic features are fed into three convolutional blocks  to capture the local patterns. Each of them has a 1D convolutional layer and a global pooling layer. To capture the temporal dependencies in the acoustic sequence, an LSTM layer and a pooling layer are employed. The obtained 4 encoded feature vectors are concatenated and fed into a fully-connected  layer to get the audio representations as follows,
\begin{equation}
\begin{array}{c}
    h_A=Concat(x_{conv_A^1},x_{conv_A^2},x_{conv_A^3},x_{lstm_A})W_A+b_A, \\
\end{array}
\label{eq:6}
\end{equation}
\begin{equation}
    x_{conv_A^i}=ConvBlock(x_A),
    \nonumber
\label{eq:4}
\end{equation}
\begin{equation}
    x_{lstm_A}=LSTMBlock(x_A),
    \nonumber
    \vspace{0.1cm}
\label{eq:5}
\end{equation}
where $ConvBlock(\cdot)=MaxPool(Relu(Conv1D(\cdot))),  \\ LSTMBlock(\cdot)=MaxPool(LSTM(\cdot)),\; x_A\in \mathbb{R}^{t_A\times d_A}$ and $ x_{conv_A^i}, \; x_{lstm_A}\in \mathbb{R}^{d_1}$ are input features and output features of convolution blocks and LSTM blocks, respectively, $W_A\in \mathbb{R}^{4d_1\times d}$ and $b_A\in \mathbb{R}^d$ are trainable parameters. For the text modality, the text features are fed into a transformer encoder of $N$ layers to capture the interactions between each pair of textual words.  An attention mechanism \cite{lian2021ctnet} is applied to the outputs of the last block to focus on informative words and generate text representations,
\begin{equation}
    h_T=(a_{fuse}^T z_T ) W_T+b_T,
\label{eq:9}
\end{equation}
\begin{equation}
    z_T=TransformerEncoder(x_T),
    \nonumber
\label{eq:7}
\end{equation}
\begin{equation}
    a_{fuse}=softmax(z_T W_z+b_z ),
    \nonumber
    \vspace{0.1cm}
\label{eq:8}
\end{equation}
where $x_T\in \mathbb{R}^{t_T\times d_T}$and $z_T\in\mathbb{R}^{t_T\times d_2}$ are input features and output features of TransformerEncoder, $W_z\in \mathbb{R}^{d_2\times 1},\; b_z\in\mathbb{R}^1,\; W_T\in \mathbb{R}^{d_2\times d}$ and $ b_T\in\mathbb{R}^d$ are trainable parameters.  $a_{fuse}\in\mathbb{R}^{t_T\times 1}$ is the attention weight.
Finally, the representations of the two modalities are concatenated and fed into three fully-connected layers with a residual connection, followed by a softmax layer. As using the label distribution to annotate emotional data, we choose the KL loss to optimize the model,
\begin{equation}
    Loss_{KL}=\sum_{i=1}^C y_i log\frac{y_i}{\hat{y_i}},
\label{eq:12}
\end{equation}
\begin{equation}
    h=Concat(h_A, h_L)W_u+b_u,
    \nonumber
\label{eq:10}
\end{equation}
\begin{equation}
\hat{y}=Softmax((h+ReLU(hW_h+b_h )) W_c+b_c ),
\nonumber
\vspace{0.1cm}
\label{eq:11}
\end{equation}
where $W_u\in R^{2d\times d},\; b_u\in \mathbb{R}^d, \; W_h\in R^{d\times d},\; b_h\in \mathbb{R}^d, \; W_c\in\mathbb{R}^{d\times c} $ and $ b_c\in\mathbb{R}^c$ are trainable parameters, $y\in\mathbb{R}^c$ is the true label distribution, $\hat{y}\in\mathbb{R}^c$  is the predicted label distribution.

\section{Experiment}
\label{sec:experiment}
\subsection{Dataset and evaluation metrics}
\label{ssec:dataset}
PDDS is evaluated on the most commonly used dataset (IEMOCAP) in SER\cite{busso2008iemocap}. There are five sessions in the dataset, where each sentence is annotated by at least three experts. Following previous work \cite{hazarika2018icon}, we choose six emotions: \textit{anger}, \textit{happiness}, \textit{sadness}, \textit{neutral}, \textit{excited} and \textit{frustration}. Session 1 to 4 are used as the training set and session 5 is used as the testing set, and PDDS is only applied on the training set. The weighted accuracy (WA, i.e., the overall accuracy) and unweighted accuracy (UA, i.e., the average accuracy over all emotion categories) are adopted as the evaluation metrics.
\subsection{Implementation Details}
\label{datails}
For audio, 512-dimensional features are extracted from the raw speech signal by the pre-trained data2vec \cite{pmlr-v162-baevski22a}. The frame size and frame shift are set to 25 ms and 20 ms, respectively. For text, the pre-trained Bert \cite{devlin-etal-2019-bert} is used to embed each word into a 768-dimensional vector.

In the original dataset, the quantity of clear samples belonging to happiness is very small, we then augment the data using mixup in order to balance the quantities of clear samples in all emotion categories. The augmentation settings are $p=0.5$, $\alpha=\beta=$ \textit{happiness}, and the quantity is increased from 43 to 215. Afterward, the data distribution smoothing is applied to the ambiguous samples of each emotion-pair with Gaussian kernel $\sigma=1$.

The kernel sizes of three CNN blocks in CLTNet are 3, 4 and 5, and the number of hidden units is 300 in LSTM. The number of layers in the transformer  encoder is 5 and each layer is with 5 attention heads. The embedding dimensions $d_1,d_2$ and $d$ in Eq. (\ref{eq:6}) and (\ref{eq:9}) are set to 300, 100 and 100, respectively. CLTNet is trained by the Adam optimizer. The learning rate,  weight decay and batch size are set to 0.001, 0.00001 and 64, respectively. The training stops if the loss does not decrease for 10 consecutive epochs, or at most 30 epochs.
\subsection{Validation experiment}
\label{ssec:val}
As PDDS is model-modality agnostic, we select SpeechText \cite{atmaja2019speech}, MAT \cite{delbrouck-etal-2020-modulated}, and MCSAN \cite{sun2021multimodal}, which aim at utterance-level emotion recognition, to evaluate its effectiveness. Our proposed CLTNet and its unimodal branches are also tested. The experimental results are shown in Table \ref{tab:1}. We can observe that all models are significantly improved when the training data are processed by PDDS, with 0.2\% $\sim$ 4.8\% increase on WA and 1.5\% $\sim$ 5.9\% increase on UA. Among them, our proposed CLTNet achieves the best performance. This result demonstrates that a reasonable data distribution in the training set does help models learn a better emotional representation.
\begin{table}[]
\centering
\caption{The comparison of models without and with PDDS.}
\setlength{\tabcolsep}{1mm}{
\begin{tabular}{@{}lcccc@{}}
\toprule
\multirow{2}{*}{Model} & \multicolumn{2}{c}{w/o PDDS} & \multicolumn{2}{c}{w/ PDDS}         \\ \cmidrule(l){2-5}
                       & WA (\%)        & UA (\%)         & WA (\%)      & UA (\%)            \\ \midrule
SpeechText \cite{atmaja2019speech}             & 55.7       & 50.6       & \textbf{58.8} & \textbf{55.1} \\
MAT \cite{delbrouck-etal-2020-modulated}                    & 54.5       & 51.0       & \textbf{58.0} & \textbf{55.5} \\
MCSAN \cite{sun2021multimodal}                  & 60.0       & 56.5       & \textbf{60.2} & \textbf{58.0} \\
CLTNet(audio only)     & 44.9       & 39.0       & \textbf{48.4} & \textbf{44.9} \\
CLTNet(text only)      & 50.9       & 45.8       & \textbf{51.6} & \textbf{50.1} \\
CLTNet                 & 55.9       & 52.3       & \textbf{60.7} & \textbf{58.2} \\ \bottomrule
\end{tabular}
}
\vspace{0.1cm}
\label{tab:1}
\end{table}

\begin{figure}[htb]
\centering
\centerline{\includegraphics[width=8.5cm]{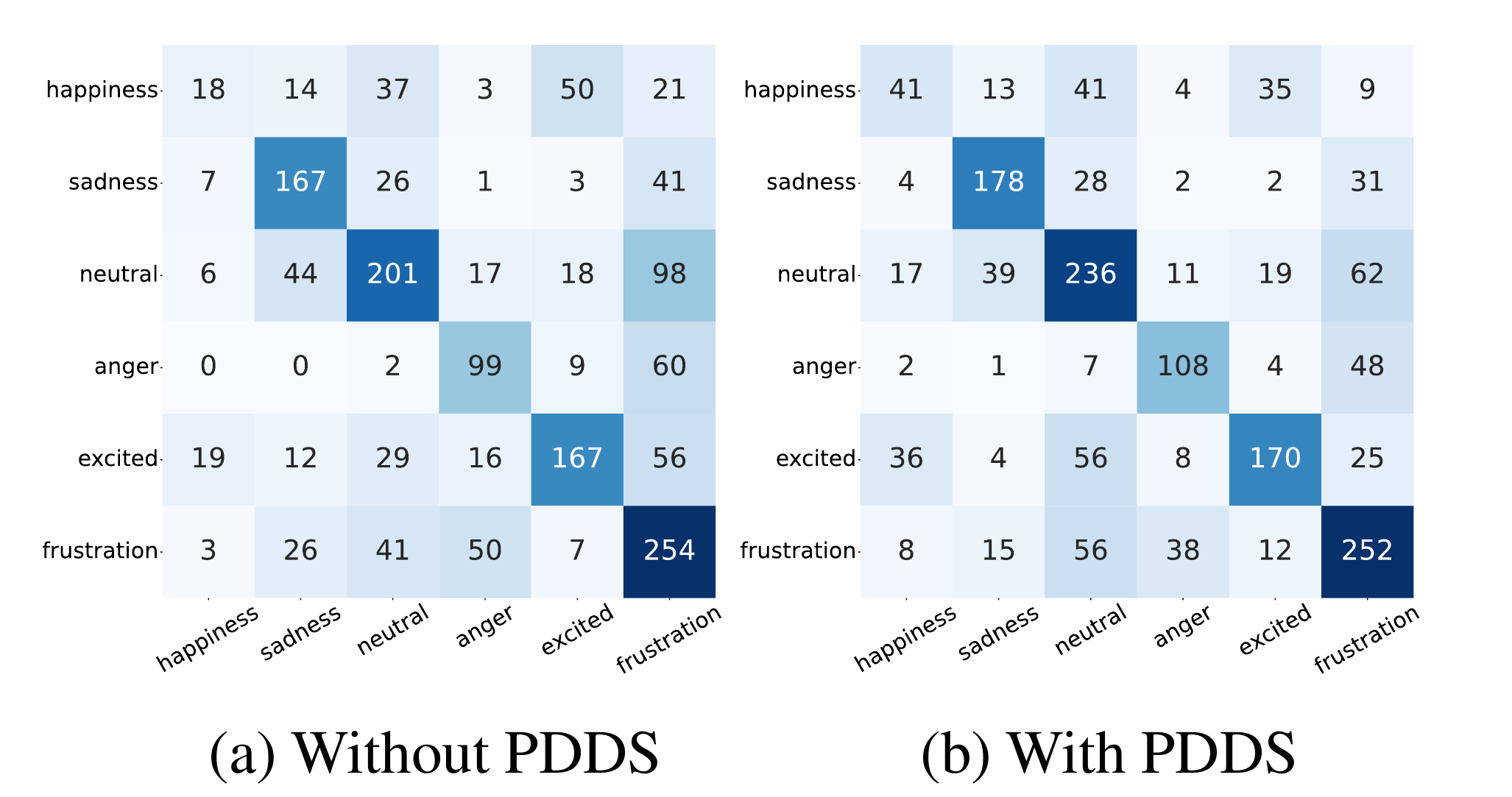}}
\caption{Confusion matrices of CLTNet without and with PDDS.}
\label{fig:5}
\end{figure}

A more detailed analysis, the confusion matrices of CLTNet with and without PDDS, are shown in Fig. \ref{fig:5}. One can see that the classification accuracies in most emotion categories are increased. Only two more samples are misclassified in \textit{frustration}. However, the rightmost columns in the confusion matrices illustrate that the model trained on the original dataset inclines to misclassify more samples as \textit{frustration}. By contrast, PDDS considerably alleviates this tendency.

\subsection{Ablation Study}
\label{ssec:ablation}
To verify the rationality of PDDS, two additional experiments are designed and tested on the CLTNet model: (a) Only balancing \textit{happiness}: Mixup augmentation is only applied on clear samples of \textit{happiness} rather than ambiguous samples. (b) Uniform balancing: the clear samples in each category and the ambiguous samples of each emotion-pair are augmented to 400 if their quantity is less than 400 (Most of them are less than 400 in the IEMOCAP dataset).

The results are shown in Table \ref{tab:2}. We can observe that all three data augmentations boost the performance of the model. Compared to only balancing the \textit{happiness} category, augmenting both ambiguous and clear samples can help the model perform better. Although Uniform balancing has the largest training dataset, PDDS performs best on WA and UA with 4.8\% and 5.9\% improvements. This result reveals the nature of advantage of PDDS is the reasonable data distribution instead of simply increasing the data size.
\begin{table}[]
\centering
\caption{Evaluation of different data augmentation methods.}
\label{tab:my-table}
\begin{tabular}{@{}lcc@{}}
\toprule
Data                                               & WA (\%)                        & UA (\%)                        \\ \midrule
Original training data                             & 55.9                           & 52.3                           \\
Only balancing \textit{happiness} & 57.4                           & 54.1                           \\
Uniform balancing                                  & 58.6                           & 56.0                           \\
With PDDS                                          & \textbf{60.7} & \textbf{58.2} \\ \bottomrule
\end{tabular}
\label{tab:2}
\end{table}

\begin{figure}[htb]
\centering
\centerline{\includegraphics[width=6.0cm]{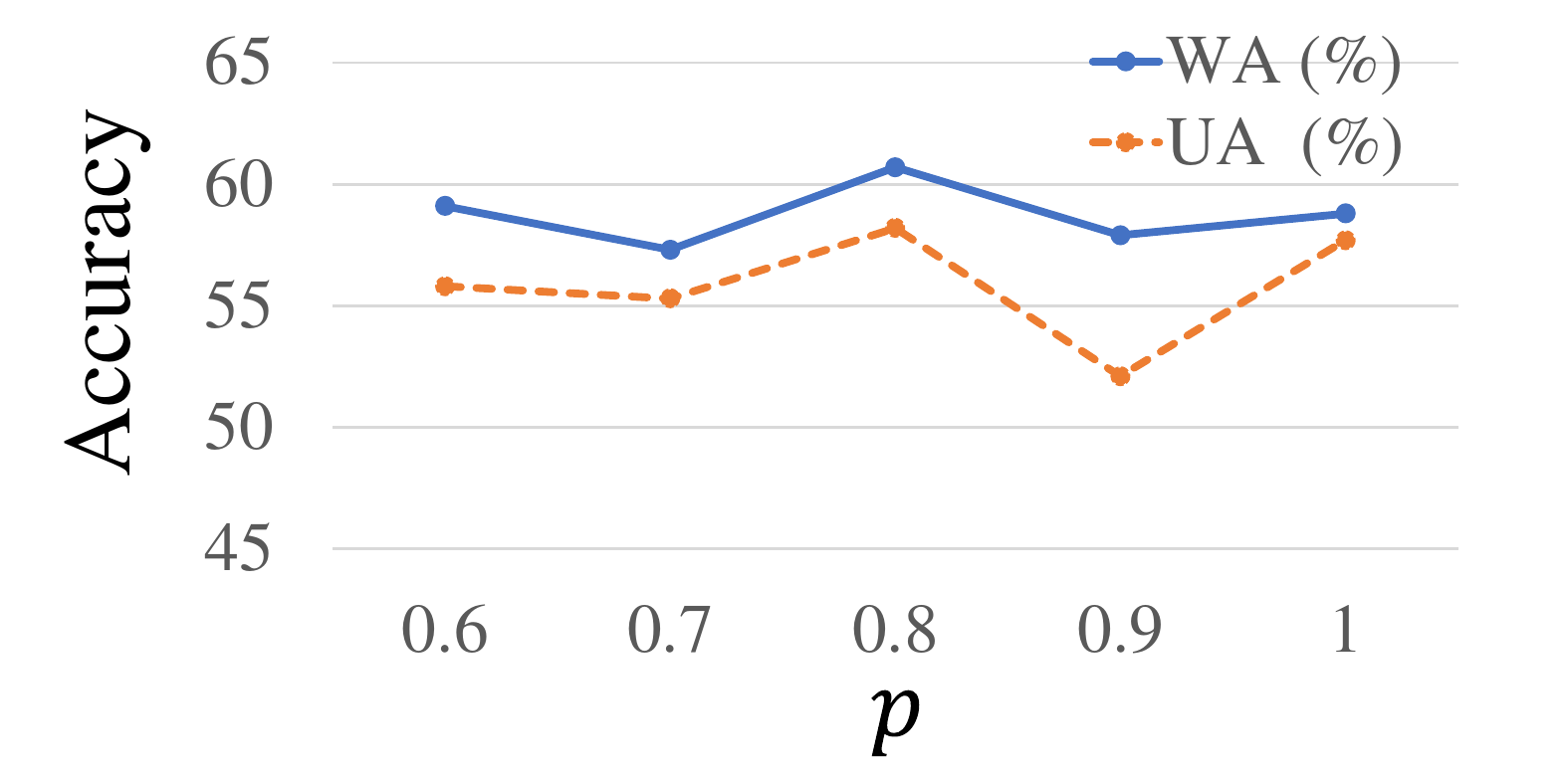}}
\caption{Effect of different $p$ values.}
\label{fig:6}
\end{figure}

We further evaluate the effectiveness of $p$ values in the mixup augmentation, and show the result in Fig. \ref{fig:6}. One can see the best results are achieved when $p=0.8$.


\section{Conclusion}
\label{sec:prior}

In this paper, we address the imbalanced data distribution in the IEMOCAP dataset by proposing the Pairwise-emotion Data Distribution Smoothing (PDDS) method. PDDS constructs a more reasonable training set with smoother distribution by applying Gaussian smoothing to emotion-pairs, and complements required data using a mixup augmentation. Experimental results show that PDDS considerably improves three SOTA methods. Meanwhile, our proposed CLTNet achieves the best result. More importantly, the ablation study verifies that the nature of superiority of PDDS is the reasonable data distribution instead of simply increasing the amount of data.  In future work, we will explore a more reasonable data distribution for a better emotional representation learning.

\vfill\pagebreak



\bibliographystyle{IEEEbib}
\bibliography{strings,refs}

\end{document}